\documentstyle[11pt,epsfig]{article}
\newcommand{\be}{\begin{equation}}
\newcommand{\ee}{\end{equation}}
\newcommand{\beqn}{\begin{eqnarray}}
\newcommand{\eeqn}{\end{eqnarray}}

\newcommand{\kf}{{\bf k}}
\newcommand{\ef}{{\bf e}}
\newcommand{\lf}{{\bf l}}
\newcommand{\Vf}{{\bf V}}
\newcommand{\flav}{\sum_f e_f^2 \alpha_{em}}
\newcommand{\Pam}{I\!\!P}
\newcounter{savefig}
\newcommand{\alphfig}{\setcounter{savefig}{\value{figure}}%
\setcounter{figure}{0}%
\renewcommand{\thefigure}{\mbox{\arabic{savefig}\alph{figure}}}}
\newcommand{\resetfig}{\setcounter{figure}{\value{savefig}}%
\renewcommand{\thefigure}{\arabic{figure}}}
\begin{document}
DESY 96-085 

ANL-HEP-PR-96-39

hep-ph/9605356     

\begin{center}
\begin{Large}
{\bf 
Azimuthal Distribution of Quark-Antiquark Jets in DIS 
Diffractive Dissociation
}\\
\end{Large}
\vspace{0.5cm}
J.\ Bartels$^a$, C.\ Ewerz$^a$, H.\ Lotter$^a$,
                        M.\ W\"usthoff\,$^b$ 
\\
\vspace{0.5cm}
$^a${II. Institut f\" ur Theoretische Physik, \\
Universit\" at Hamburg, Germany}\footnote
        {\noindent
        Supported by Bundesministerium f\"ur Forschung und
        Technologie, Bonn, Germany under Contract 05\,6HH93P(5) and
        EEC Program "Human Capital and Mobility" through Network
        "Physics at High Energy Colliders" under Contract
        CHRX-CT93-0357 (DG12 COMA).}
\\
$^b${High Energy Physics Division, \\
Argonne National Laboratory, USA}\footnote{\noindent
Supported by the US Department of Energy, High Energy Physics Division,
Contract W-31-109-ENG-38.}
\end{center}
\vspace{2.0cm}
\begin{abstract} 
\noindent
We investigate the azimuthal distribution of  
quark-antiquark 
jets in DIS diffractive dissociation  
with large transverse momentum.
In this kinematical region the matrix element
is expressed in terms of the gluon structure function.
For the transverse part of the cross section we find a 
$\cos(2 \phi)$-distribution with the maximum at 
$\phi = \pm \pi/2$, i.e.\ the jets prefer a direction 
perpendicular to the electron plane. 
This is in contrast to boson gluon fusion where 
the $q\bar{q}$ jet cross section
for transversely polarized bosons peaks at $\phi=0$ and $\phi=\pi$. 
We discuss the origin of this striking difference 
and present numerical results relevant for the diffractive dissociation 
at HERA.
\end{abstract}
\vspace{2cm}
{\bf 1.)}
The observation and analysis of diffractive events in deep inelastic electron
proton scattering at HERA \cite{H11,H12,ZEUS1,ZEUS2} has attracted 
much attention recently. Particular interest has been given to the question
whether this process is dominated by the same (soft) Pomeron which  
is observed in diffractive dissociation of hadron-hadron scattering, or
by the hard Pomeron which is associated with the observed rise of $F_2$ at
small $x$. As far as the (inclusive) diffractive 
structure function is concerned,
one expects to see a combination of both the soft and the hard Pomeron.
As a method of separating the hard component it has recently been suggested
to look for final states of the diffractive system which consist of jets with
large transverse momenta. The simplest configuration is a two-jet final state 
of a quark-antiquark pair. The cross section for this process, together
with a few numerical predictions, has been presented in \cite{blw}. 
In this letter we continue this analysis by calculating, in the 
$\gamma^* - \mbox{Pomeron}$
center of mass system, the azimuthal distribution
of the jets relative to the plane formed by the beam axis and the scattered
electron. As we will show, this distribution differs from that of
the photon-gluon fusion process and therefore presents a characteristic signal
of the two gluon model for the hard Pomeron.
\\ \\
{\bf 2.)}
The process that we are going to investigate is shown in Fig.\ \ref{f1a},
and the variables are illustrated in Fig.\ \ref{f1b}; 
for simplicity we restrict 
ourselves to the forward direction $t=0$. Throughout this paper we work in the
Pomeron-photon center of mass system. We begin with the subprocess
$\gamma^* + \mbox{proton} \to q\bar{q} +\mbox{proton}$, 
and we take the incoming photon to
be linearly polarized in the transverse direction. 
The polarization vectors are $e_x=(0,1,0,0)$ and 
$e_y=(0,0,1,0)$, and the electron momenta lie in the $x$-direction.
We work in the leading-log $(1/x)$ approximation and 
use the $\kf$-factorization theorem \cite{cat} to express 
the amplitude through the unintegrated gluon distribution of the
proton. 
This is based on the assumption that the total energy $s=(q+p)^2$
is much larger than the photon virtuality $Q^2=-q^2$ and the invariant
mass of the outgoing quark anti-quark pair $M^2=(q+x_{\Pam}p)^2$.
The quark phase space is parameterized in terms of the quarks 
transverse momentum $k_{\perp}=|k_{\perp}|(e_x \cos \phi +
e_y \sin \phi)$, and its longitudinal  
momentum fraction $\alpha$ according to the Sudakov decomposition
$k_{\mu}=\alpha q'_{\mu} + \kf^2/(2 \alpha  p q') p_{\mu} + k_{\perp \mu}$
with $q'=q+xp$ and $\kf^2 = - k_{\perp \mu}k^{\mu}_{\perp}$.
The invariant mass $M^2$ is related to the transverse momentum
through $\kf^2=\alpha(1-\alpha)M^2$.
The differential cross section for the 
incoming photon with a polarization lying in the electron plane has the form:
\beqn
\frac{d^{\gamma^* p}_{D,T}}{dM^2 d \kf^2 dt d\phi}_{|t=0} =
\frac{d^{\gamma^* p}_{D,T}}{dM^2 d \kf^2 dt}_{|t=0}
 - 2 \frac{\frac{\kf^2}{M^2}}{1-2\frac{\kf^2}{M^2}} \cos 2\phi
\frac{d^{\gamma^* p}_{D,T}}{dM^2 d \kf^2 dt}_{|t=0}
\label{e1}
\eeqn
where $\frac{d \sigma_{D,T}^{\gamma^* p}}{dM^2d \kf^2 dt}$ will be given below
in eq.\ (\ref{gampt}). 
The main feature to be noticed is the minus sign in front of the
angular dependent piece of the cross section. It is opposite to the sign
structure of the photon-gluon fusion process \cite{lit} (which will be
discussecd below), and it leads to a
characteristic maximum of the cross section\footnote{\noindent
A similar phenomenon has been observed in the azimuthal distribution
of forward jets in deep inelastic electron proton scattering
\cite{BDDGW,BDW}.} at $\phi= \pi /2$. 
For the incoming photon being polarized in the
$y$-direction the cross section has the same form as (1), but with a plus
sign in front of the second term.\\ \\
Turning to electroproduction, we have to sum
over all polarizations of
the virtual photon; the interference of 
longitudinal and transverse polarizations 
leads to an additional azimuthal
dependence. We need the Sudakov parametrization of the electron momentum:
\beqn
\ell_{\mu} =  \frac{1}{y} q'_{\mu} + 
           (1-y) \frac{x}{y} p_{\mu} + \ell_{\perp \mu}
\;\;, \;\; \ell_{\perp}^2= - \frac{1-y}{y^2}\,Q^2
\;\;, \;\; \ell_{\perp}= |\ell_{\perp}| e_x
\eeqn
Inserting this into the usual lepton tensor:
\beqn
L_{\mu \nu} = 2 \left[\ell_{\mu}\ell_{\nu}- \frac{Q^2}{4} g_{\mu \nu}
\right]
\eeqn
and contracting with the photon polarization vectors
we arrive at the following flux factors for the linearly polarized transverse
photons
\beqn
L_{\mu\nu} e_x^{\mu} e_x^{\nu}&=&\frac{1}{2}\, Q^2 + 2 \frac{1-y}{y^2}\,Q^2
\label{flux}
\\
L_{\mu\nu} e_y^{\mu} e_y^{\nu}&=&\frac{1}{2}\, Q^2  
\eeqn
Summing over all polarizations we arrive at the $ep$-cross section:
\beqn
\frac{ d \sigma^{e^- p}_D}
{d y d Q^2 d M^2 d \kf^2 d \phi d t}_{|t=0}
= \frac{\alpha_{em}}{y Q^2 \pi}
\left[
\frac{1+(1-y)^2}{2}
\frac{d \sigma^{\gamma^* p}_{D,T}}{d M^2 d \kf^2 d t}_{|t=0}
- 2 (1-y) \frac{\frac{\kf^2 }{M^2}}{1 - 2 \frac{\kf^2}{M^2}} \cos 2 \phi
\frac{d \sigma^{\gamma^* p}_{D,T}}{d M^2 d \kf^2 d t}_{|t=0}\right.
\nonumber \\ 
\left. + (1-y) \frac{d \sigma^{\gamma^* p}_{D,L}}{d M^2 d \kf^2 d t}_{|t=0}
+(2-y)\sqrt{1-y} \cos \phi
\frac{d \sigma^{\gamma^* p}_{D,I}}{d M^2 d \kf^2 d t}_{|t=0}
\right] \;\;\;\;\;\;
\label{eq1}
\eeqn
where the indices T, L, and I refer to the contributions of transverse 
and longitudinal photons and the interference term, resp. Apart from the 
flux factors
of the photons, the transverse part has the same structure as seen in 
eq.\ (\ref{e1}).
In particular, the minus sign in front of the angular dependent term is a 
direct consequence of combining (\ref{e1}) with the flux factor in 
(\ref{flux}).
\\ \\
Eq.(6) is written for the case in which the
polar-angle $\theta$ ($\cos \theta=1-2\alpha$),
the angle between the quark jet with momentum $k$ and the proton (Pomeron), 
is restricted to be smaller than $\pi/2$, i.e. $\alpha$
varies between 0 and 1/2. If this jet lies in the other hemisphere (polar
angle $\theta$ between $\pi/2$ and $\pi$ or, equivalently, $\alpha$ between
$1/2$ and $1$), the last term in (6) changes sign. 
When averaging over both hemispheres the
cross section above has to be multiplied with 
a factor of two, and the interference term disappears. 
Then, the only angular dependence results from the
term proportional to $\cos 2 \phi$ in eq.\ (\ref{eq1}). 
Note the additional factor $\kf^2$ in front of this term. We therefore 
expect the azimuthal asymmetry to be 
more pronounced at large transverse momenta (at the same time, however, the
cross section decreases with growing transverse momentum). 
Finally, by integrating the cross section from $0$ to $2 \pi$ 
we recover our previous result \cite{blw}.   
A numerical analysis of the above formulae will be performed in part 
{\bf 4.)}. \\ \\
The expressions for the 
longitudinal and transverse
photon-proton
cross sections have been obtained in \cite{blw}:
\beqn
2 \pi \frac{d \sigma^{\gamma^* p}_{D,L}}{d M^2 d \kf^2 d t}_{|t=0}
&=&
\frac{1}{M^4} \frac{\kf^2}{Q^2} \frac{1}{\sqrt{1-4 \frac{\kf^2}{M^2}}}
\frac{2}{3}
\flav \pi^2 \alpha_s^2 \left[ I_L(Q^2,M^2,\kf^2) \right]^2
\\
2 \pi \frac{d \sigma^{\gamma^* p}_{D,T}}{d M^2 d \kf^2 d t}_{|t=0}
&=&
\frac{1}{M^4} 
\frac{1-2 \frac{\kf^2}{M^2}}{\sqrt{1-4\frac{\kf^2}{M^2}}}
\frac{1}{24}
\flav \pi^2 \alpha_s^2 \left[ I_T(Q^2,M^2,\kf^2) \right]^2
\label{gampt}
\eeqn
In the same spirit we find for the interference contribution
\beqn
2 \pi \frac{d \sigma^{\gamma^* p}_{D,I}}{d M^2 d \kf^2 d t}_{|t=0}
&=&
\frac{1}{M^4} \left(\frac{\kf^2}{Q^2}\right)^{\frac{1}{2}}
\frac{1}{6}
\flav \pi^2 \alpha_s^2   
I_T(Q^2,M^2,\kf^2) I_L(Q^2,M^2,\kf^2) 
\eeqn
with
\beqn
I_T &=& - \int  \frac{d \lf^2}{\lf^2} 
\left[ \frac{M^2-Q^2}{M^2+Q^2}+\frac{\lf^2+\frac{\kf^2}{M^2}(Q^2-M^2)}
{\sqrt{(\lf^2+\frac{\kf^2}{M^2}(Q^2-M^2))^2+4 \kf^4 \frac{Q^2}{M^2}}}
\right]
{\cal F}_G(x_{\Pam},\lf^2)
\\
I_L &=& - \int  \frac{d \lf^2}{\lf^2} 
\left[
\frac{Q^2}{M^2+Q^2} - \frac{\kf^2 Q^2}
{M^2 \sqrt{(\lf^2+\frac{\kf^2}{M^2}(Q^2-M^2))^2+4 \kf^4 \frac{Q^2}{M^2}}}
\right]
{\cal F}_G(x_{\Pam},\lf^2)
\eeqn
and the unintegrated gluon distribution of the proton
\beqn
\int^{Q^2}d \lf^2 {\cal F}_G(x_{\Pam},\lf^2)=x_{\Pam}G(x_{\Pam},Q^2)
\label{bfkl}
\eeqn

%
\noindent
In our expression (\ref{bfkl}) we have used, for the two-gluon subamplitude,
the (forward) gluon structure function of the proton. Strictly speaking,
this is not quite accurate, because the longitudinal momenta of the two
gluon lines differ by $x_{I\!\!P} p$. In our leading-log calculation, however,
we cannot distinguish between this 'slightly nonforward' gluon structure
function and the usual DIS gluon structure function. We therefore 
shall use the usual gluon density of the proton, and, as a consequence,
we have to accept an error in the absolute normalization which is 
characteristic for a leading-log calculation (cf.\ the discussion in 
\cite{blw}). \\ \\
Without going into more detail we quote the results that were 
obtained in \cite{blw} for the functions $I_T,I_L$ above
\beqn
I_T &=& \left[\frac{4 Q^2 M^4}{\kf^2(M^2+Q^2)^3} + b_T 
\frac{\partial}{\partial \kf^2}
\right] x_{\Pam}G(x_{\Pam},\kf^2 \frac{Q^2+M^2}{M^2}) 
\label{it}
\\
I_L &=& \left[\frac{ Q^2 M^2 (Q^2-M^2)}{\kf^2(M^2+Q^2)^3} + b_L 
\frac{\partial}{\partial \kf^2}
\right] x_{\Pam}G(x_{\Pam},\kf^2 \frac{Q^2+M^2}{M^2})
\label{il}
\eeqn
In these expressions the first terms represent the double leading log 
approximation. As discussed in \cite{blw}, the 
derivative terms with the functions $b_T,b_L$ 
of $Q^2$ and $M^2$ are those next-to-leading order 
(in $\log \kf^2(M^2+Q^2)/M^2$) corrections which, as we believe, are
numerically most important. They can be found in
eqs.\ (12), (13) of \cite{blw} and will not be given explicitly here.
We wish, however, to stress that these next-to-leading order corrections 
are not
complete, and we believe that a systematic study of order-$\alpha_s$ 
corrections to our 'Born approximation' remains an important future task.
\\ \\ 
{\bf 3.)}
It will be useful to compare our results with the usual photon gluon fusion
cross section. To be definite, let us, again, consider the 
cross section for the production of a quark-antiquark pair in the kinematic
region where $W^2$ is much larger than $Q^2$. The momenta are labelled as
shown in Fig.\ \ref{f2}. 
The gluon momentum $l_{\mu}$ has the Sudakov decomposition
$l=\eta p + l_{\perp}$ 
with $\eta=x_B(1+W^2/Q^2)$  
(the component along the momentum $q'$ is small
and can be neglected), and the polarization vector of the gluon is
$p_{\mu}\sqrt{2/W^2}$. 
Choosing the incoming photon to have the transverse
polarization in the $x$-direction, and summing over the quark helicities,
we obtain for the square of the subprocess $\gamma^*+g \to q \bar{q}$:
\beqn
\ef_x \cdot \ef_x \,\, \Vf \cdot \Vf -
\,
4\alpha(1-\alpha)
\ef_x \cdot \Vf \;\; \ef_x \cdot \Vf
\label{str}
\eeqn
where we have disregarded overall constants, and the vector 
$\Vf$ is defined as
\beqn
\Vf= \frac{\kf}{D(\kf)}
-\frac{\kf-\lf}{D(\kf-\lf)}
\;\;
, \,\, 
D(\kf)=\alpha (1-\alpha) Q^2 +\kf^2
\label{wf}
\eeqn
As can be seen from (\ref{wf}), the vector $\Vf$ can be interpreted as the
$q \bar{q}$ component of the wave function of the transverse photon. It has
the form of a dipole.
The standard result for the photon-gluon 
fusion process is obtained by integrating
over the azimuthal angle $\phi_l$ of the gluon and taking the limit 
$\lf^2 \to 0$. For small $\lf^2$ one finds
\beqn
\int \frac{d \phi_l}{2 \pi} \Vf_i \Vf_j = \frac{\lf^2}{2 D(\kf)^2}
\left(
\delta_{ij} - \kf_i \kf_j \frac{\kf^2 Q^2}{M^2 [D(\kf)]^2}
\right)
\label{lim1}
\eeqn
and the cross section becomes
\beqn
d \sigma \sim \left( 1 -2 \frac{\kf^2}{M^2} \right) 
  \frac{Q^4 + M^4}{(Q^2+M^2)^2}
+ 4 \frac{\kf^2}{M^2} \cos 2\phi \frac{Q^2 M^2}{(Q^2 + M^2)^2}
\label{lim2}
\eeqn
The angular part now has a positive sign, quite in contrast to the
previous case ((\ref{e1}) or (\ref{eq1})). 
This difference can be traced back to the 
$\delta_{ij}$ piece
in eq.\ (\ref{lim1}): the second term 
$\sim \kf_i \kf_j$ alone would have lead to the same 
structure as (\ref{e1}), 
and it is the first term which leads to the sign change in
(\ref{lim2}). 
As an attempt to find a physical interpretation, one might interpret
the contribution $\sim \kf_i \kf_j$ as an incoherent product of the two
dipoles, whereas the term $\sim \delta_{ij}$ represents a correlation between
the dipole and its conjugate which results from the angular integral. \\ \\
For comparison we write down the analogue of 
eq.\ (\ref{str}) for the diffractive case, 
i.e.\ the expression for the square of the subprocess
$\gamma^* + 2g \to q \bar{q}$ (Fig.\ \ref{f1a}). It has the same form as 
(\ref{str}), but now 
we have two $\lf$-integrals (both for the matrix element and its complex 
conjugate). The wave functions $\Vf$ in (\ref{str}) have to be replaced 
by $\Vf_D(\lf)$ and $\Vf_D(\lf')$:
\beqn
\ef_x \cdot \ef_x \,\, \Vf_D (\lf) \cdot \Vf_D (\lf') -
\, 4\alpha(1-\alpha)\,
     \ef_x \cdot \Vf_D (\lf) \;\; \ef_x \cdot \Vf_D (\lf')
\eeqn
where
\beqn
\Vf_D (\lf) =2 \frac{\kf}{D(\kf)} - \frac{\kf-\lf}{D(\kf-\lf)} - 
            \frac{\kf+\lf}{D(\kf+\lf)}.
\eeqn 
The limit $\lf^2 \to 0$ and the angular integration are done independently for
both $\lf$ and $\lf'$. 
For the leading term in this limit we find
\beqn
\int \frac{d \phi_l}{2 \pi} \Vf_D (\lf)=
\kf \frac{4 \,Q^2 M^4}{(Q^2+M^2)^3}\frac{\lf^2}{\kf^4}
\left[1 + O(\lf^2)\right]
\eeqn
As a result, the tensor $\Vf_i \Vf_j$ is proportional to $\kf_i \kf_j$, and
there is no contribution proportional to $\delta_{ij}$.\\ \\
To complete our review of the photon-gluon fusion process, we mention
that - unlike in the usual discussion where only the leading term of the
limit $\lf^2 \to 0$ is retained - a consistent application of the 
$\kf_T$-factorization allows to consider also the case $\lf^2 \neq 0$. 
As one can see 
immediately from the kinematics in Fig.\ \ref{f2}, in this case the momentum
transfer from the proton system is no longer zero. 
When integrating over $\lf$, one at the same time
averages over the transverse momentum of one of the two jets, and
as a result the cross section can no longer be compared with the diffractive
cross section of the first part.
Nevertheless, we still can ask for the azimuthal dependence of the jet with 
fixed momentum $\kf^2$. 
Having performed the $\phi_l$ integral, one arrives at a tensor
structure similar to (\ref{lim1}), 
but the coefficient functions of the two terms
might be quite different. Consequently, also the sign structure in the
cross section in (\ref{lim2}), 
which results from a combination of these coefficient 
functions,
may change, and the azimuthal dependence of the cross section will 
depend upon the $\lf$-dependence of the gluon distribution. A definite 
prediction could be made if one uses the BFKL Pomeron as a model for
the gluon structure function. In analogy with the results of \cite{blw} one
expects, as a correction to the DLA result (\ref{lim2}), 
a term containing the 
unintegrated gluon distribution. Details of such a calculation will be
presented elsewhere \cite{l}.\\ \\  
{\bf 4.}
In this section we want to present some numerical results
based on the formula eq.\ (\ref{eq1}).
In our analysis we closely follow the strategy described
in \cite{blw}. 
For the diffractive cross section the integration
over $t$ is performed after multiplication
of our $t=0$ expression   
with the Donnachie--Landshoff formfactor. 
We use the GRV next to leading order parameterization \cite{grv}
of the gluon density and take account of the 
subleading corrections, indicated as $b_T, b_L$ in eqs.\ 
(\ref{it}), (\ref{il}).
For the nondiffractive case where we do not calculate 
corrections of this kind, we take the GRV
leading order density. 
In both cases we take $\alpha_s$ running with the scale 
equal to the momentum scale of the gluon density. 
\\ \\
First, to get an overall impression we show in Fig.\ \ref{f3},
as a function of the azimuthal angle $\phi$, the totally 
integrated electron proton cross section with 
the kinematical constraints $Q^2 > 10 \, \mbox{GeV}^2, x_{\Pam}
< 10^{-2}, \kf^2 > 2 \,  \mbox{GeV}^2$ and $50 \, \mbox{GeV}^2
< W < 220 \, \mbox{GeV}^2$. Note that, for given values of $M^2$ and
$\kf^2$, we have summed over the two configurations with 
$\alpha=\frac{1}{2}+\frac{1}{2}\sqrt{1-4\kf^2/M^2}$ and
$\alpha=\frac{1}{2}-\frac{1}{2}\sqrt{1-4\kf^2/M^2}$, 
i.\ e.\ over 
the two different hemispheres. 
In addition we added the degenerate contribution with 
$\phi+\pi$, i.\ e.\ the total cross section is recovered by integrating
$d \sigma / d \phi$ from $\phi=0$ up to $\phi=\pi$.
As a result, the interfence term in (6)
drops out, and our curves are symmetric w.r.t. $\phi=\pi/2$.
For comparison we also show the contribution
of the longitudinal polarization (dashed line) and the $\phi$-independent 
part of the transverse polarization (dotted line). The angle dependent 
contribution leads to a factor of 
approximately 2 between the minimum of the cross 
section at $\phi=0,\pi$ and the maximum at $\phi=\pi/2$. Furthermore, 
one notices that the longitudinal polarization gives on the whole a 10\% 
contribution to the total cross section. As to the differential cross 
section, the relative magnitude of the maximum, as well as the contribution 
of the longitudinal polarization, will depend on the kinematic variables,
especially $\beta = Q^2/(Q^2 + M^2)$ (cf.the discussion in ~\cite{blw}). 
\\ \\
Next we want to exhibit the influence of the kinematical prefactors 
on the relative magnitude of the azimuthally asymmetric contribution.
In Fig.\ \ref{f4} we show again the totally integrated $ep$ - cross
section with the above 
given cuts for $Q^2,x_{\Pam}$ and $W$ but with
three different cuts in $\kf^2$. The curves demonstrate that a restriction
of the phase space to larger $\kf^2$ leads to an increase of the asymmetry,
due to an additional $\kf^2$ in the prefactor of the angular dependent 
term. The absolute magnitude of the cross section, however, decreases
due to the overall suppression at large transverse momenta.
\\ \\
Instead of $\kf^2$, it might be more convenient to use the polar angle
$\theta$ (Fig.\ 2) between the jet with momentum $k$ and the Pomeron momentum
in the photon-Pomeron cm-system.
This angle is determined through $\sin^2 \theta = 4 \kf^2 /M^2$.
Here we do not add the configuration at $\phi+\pi$.
The cross section for $\phi > \pi$ is obtained by reflection
w.r.t.\ the axis $\phi=\pi$.   
To make sure that we do not get a contribution from a region where
our perturbative analysis is not reliable we have imposed an absolute
lower cutoff on $\kf^2$ of $1 \, \mbox{GeV}^2$. 
In addition we have chosen bins in $M^2$ to keep the center of mass
energy of the photon-pomeron subprocess roughly constant.
In Fig.\ \ref{f5} we present the cross section with cuts in 
$\theta$ instead of cuts in $\kf^2$ imposed. Now the interference term in
(6) can no longer be neglected, and we expect a slight asymmetry between
$\phi=0$ and $\phi=\pi$.
In the left diagram we have chosen $20 \,\mbox{GeV}^2 < M^2 < 
50 \,\mbox{GeV}^2$ and two different cuts in $\theta$, 
$\pi/16 < \theta < \pi/4$ 
(upper curve) and $\pi/4 < \theta < \pi/2$
(lower curve), respectively.
Again, as expected, for larger $\theta$,
corresponding to larger $\kf^2$, the asymmetry
is more pronounced, but the normalization is smaller.
The $\cos \phi$-asymmetry is clearly visible and it can be seen that
the coefficent of the $\cos \phi$-term is positive.
In the right diagramm we have chosen the same $\theta$-cuts
but the $M^2$-bin $50 \,\mbox{GeV}^2 < M^2 < 
100 \,\mbox{GeV}^2$. Here the coefficent of the $\cos \phi$-term is negative.
This is due to the fact that for large masses the function $I_L$
(eq.\ (\ref{il})) becomes negative. 
This sign change is a unique property of the interference term.
\\ \\
Finally we want to illustrate the essential difference
between the diffractive cross section based on
the two gluon exchange model and the boson gluon fusion
cross section. 
We have calculated the total cross section for quark-antiquark
jet production in $ep$-scattering based on the photon gluon 
fusion mechanism. The explicit formulae for this cross 
section can be obtained with the method sketched in {\bf 3.)}
and can be found in \cite{lit}. In Fig.\ \ref{f6} we present
the normalized (to unit integral) 
differential cross section in $\phi$ with 
all other variables integrated using the cuts 
$Q^2 > 10 \, \mbox{GeV}^2, \eta
< 10^{-2}, \kf^2 > 2 \,  \mbox{GeV}^2$ and $50 \, \mbox{GeV}^2
< W^2 < 220 \, \mbox{GeV}^2$.   
This curve nicely demonstrates that 
in the boson gluon fusion case
the jets prefer 
a direction in the electron plane (defined by $\phi=0$),
whereas in the two gluon 
exchange case (Figs.\ \ref{f3}-\ref{f5}) the jets prefer 
a perpendicular direction.
\\ \\
{\bf 5.)}
In this paper we have calculated the azimuthal dependence of two 
quark-antiquark jets with large transverse momenta in DIS
diffractive dissociation. Our main result, contained in 
eq.\ (\ref{eq1}), is the
striking sign structure of the angle dependent term. It leads to a 
characteristic maximum of the cross section at $\phi=\pi/2$, 
quite in contrast
to the angular dependence of the photon-gluon fusion subprocess which peaks
at $0$ and $\pi$. An experimental observation of this 
maximum at $\pi/2$ 
would clearly confirm our present understanding of the
'hard Pomeron'. \\ \\
Some time ago it has been suggested that the photon-gluon fusion subprocess,
accompanied by the additional exchange of soft gluons \cite{BH} or by
a suitable modification of the final state interactions \cite{IR} might
represent the basic mechanism of the diffractive dissociation observed at
HERA. In principle, the different dependencies upon $\phi$ discussed in this
letter might help to discriminate between the two mechanisms: the exchange
of a color singlet two-gluon state (hard Pomeron) and the single gluon
exchange in the photon-gluon fusion model. There are, however, two 
{\it caveats}
to be kept in mind. First, the perturbative calculation of the two-gluon 
exchange can be justified only for jets with large transverse momenta; 
this excludes any prediction for the diffractive structure function which 
may very well be dominated by
soft final states (e.g.\ Pomeron-remnant jets). The discussion in \cite{BH},
on the other hand, refers to the diffractive structure function and not to
specific final states. Secondly, for the case of the
photon-gluon fusion we do not know to what extent final state hadronization
will modify the azimuthal dependence of the partonic cross section: since the
$q\bar{q}$ pair is in a color octet state, one expects a color connection
with the proton system which may affect the angular distribution. In contrast,
for the two-gluon exchange model we believe that the partonic cross section
is rather robust against hadronization effects. Namely,
our perturbative calculation describes the azimuthal distribution
of the $q\bar{q}$ dipole which initially has a small transverse size.
The subsequent hadronization will be restricted to the $q\bar{q}$-system,
and thereby not change the orientiation.\\ \\
In summary, final states in the diffractive dissociation of the photon 
which contain
only hard jets represent a novel class of hard processes which are calculable
in perturbative QCD: jet production from the `annihilation' of a photon and
a color singlet two-gluon state (a hard Pomeron). The $q \bar{q}$ jets 
considered in \cite{blw} and
in the present paper represent the very first step along this line  -- 
somewhat analogous to the two-jet final states in $e^+ e^-$ annihilation. 
What has to come next is the order-$\alpha_s$ correction to the two-jet 
configuration, together with the three-jet final state.\\ \\
%

\section*{Figure Captions}
\begin{description}
\item
Fig.\ \ref{f1a} :
One of the four diagrams contributing to the hard 
           process $\gamma^* + p \rightarrow q\bar{q} + p$. 
           The outgoing (anti)quark momenta are fixed and
           the same for all four diagrams. The blob depicts 
           the unintegrated gluon structure function ${\cal F}_G$.
\\
\item
Fig.\ \ref{f1b} :
Definition of planes and angles in the 
           $\gamma^* I\!\!P$ cms ($\vec{q}+x_{I\!\!P}\vec{p}=0$).
           The leptonic plane is given by the vectors
           $\vec{\ell}$ and $\vec{q}=\vec{\ell}-\vec{\ell}^\prime$, 
           the jet plane by the vectors $x_{I\!\!P}\vec{p}$
           and $\vec{k}$). $\phi$ is the angle between these two
           planes. $\theta$ is the polar angle of the jet direction 
           $\vec{k}$.
\\
\item
Fig.\ \ref{f2} :
           One of the two diagrams for the boson gluon fusion 
           (BGF) process.
           The square of the BGF amplitude again leads to the 
           gluon structure function.
\\
\item
Fig.\ \ref{f3} :
The $\phi$ dependence of the $ep$ - cross section in eq.\ (\ref{eq1})
integrated over the other variables. Displayed is the sum of all
terms (solid line), and the contribution of the angular independent
transverse (dotted line) and longitudinal (dashed line) terms.  
\\
\item
Fig.\ \ref{f4} :
The $\phi$ dependence of total $ep$ - cross section 
with different cuts in $\kf^2$: 
$1 \, \mbox{GeV}^2 < \kf^2 < 2 \, \mbox{GeV}^2$,
$2 \, \mbox{GeV}^2 < \kf^2 $,
$5 \, \mbox{GeV}^2 < \kf^2 $ (from top to bottom). 
\\
\item
Fig.\ \ref{f5} :
The $\phi$ dependence of total $ep$ - cross section 
with different cuts in $\theta$ and $M^2$. 
In the left diagram we have 
$20 \, \mbox{GeV}^2 < M^2 < 50 \, \mbox{GeV}^2$ 
and $\pi/16 < \theta < \pi/4$ (upper curve),
$\pi/4 < \theta < \pi/2$ (lower curve).
In the right diagram we have 
$50 \, \mbox{GeV}^2 < M^2 < 100 \, \mbox{GeV}^2$ 
and $\pi/16 < \theta < \pi/4$ (upper curve),
$\pi/4 < \theta < \pi/2$ (lower curve).
\\
\item
Fig.\ \ref{f6} :
Normalized differential cross section for two jet production
based on the boson-gluon fusion mechanism.
\end{description}
\section*{Figures}
\setcounter{figure}{1}
\alphfig
\begin{figure}[htbp]
\begin{center}
\input diffkin.pstex_t
\end{center}
\caption{
\label{f1a}}
\end{figure}
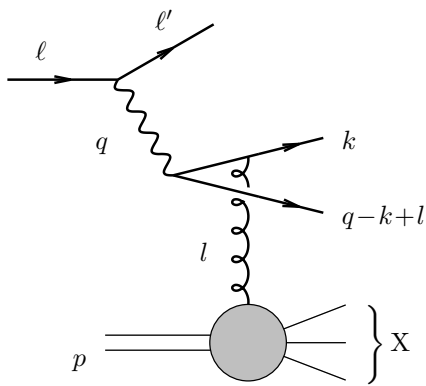
\begin{figure}[htbp]
\begin{center}
\input verkleiner.pstex_t
\end{center}
\caption{
\label{f1b}}
\end{figure}
\resetfig
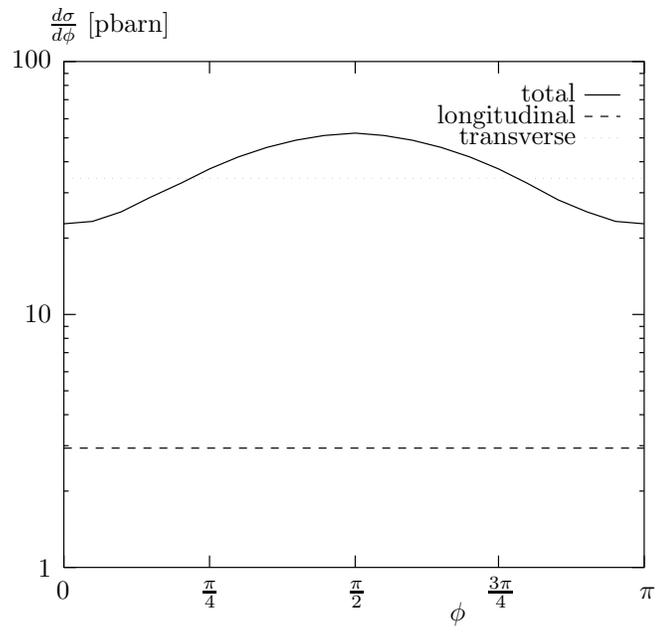
\begin{figure}[htbp]
\begin{center}
\input bgfkin.pstex_t
\end{center}
\caption{
\label{f2}}
\end{figure}
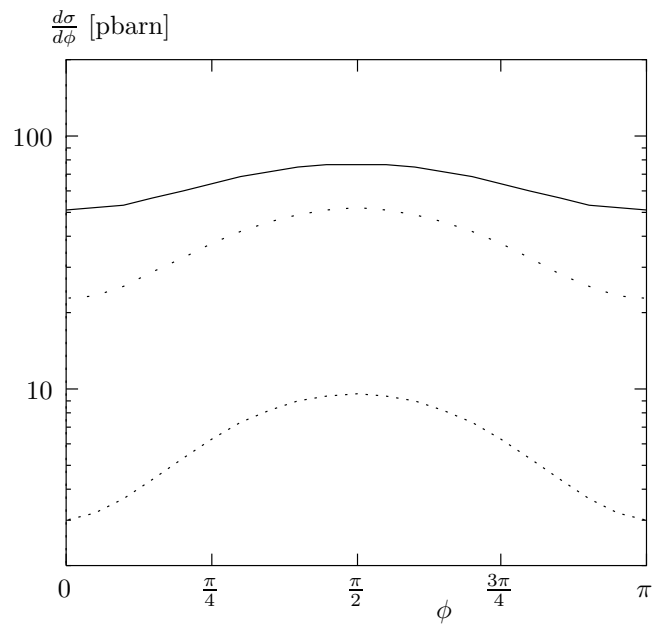
\begin{figure}[htbp]
\begin{center}
\input fig1.pstex_t
\end{center}
\caption{
\label{f3}}
\end{figure}
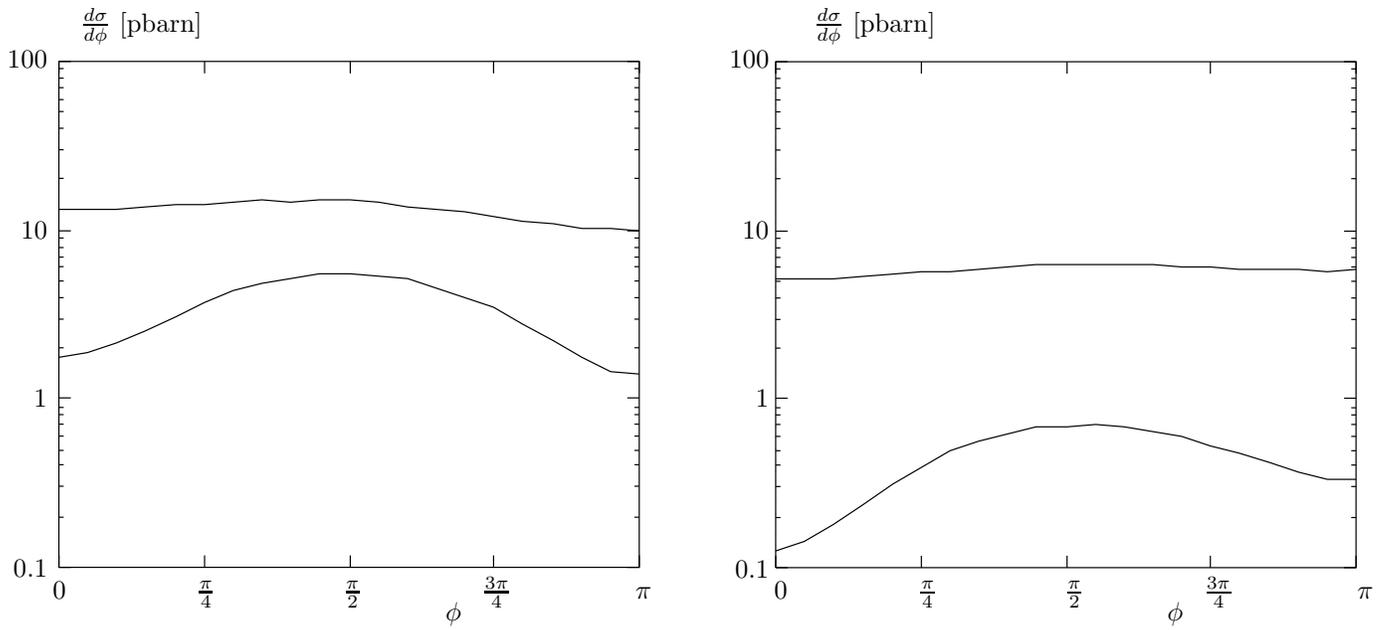
\begin{figure}
\begin{center}
\input fig2.pstex_t
\end{center}
\caption{
\label{f4}}
\end{figure}
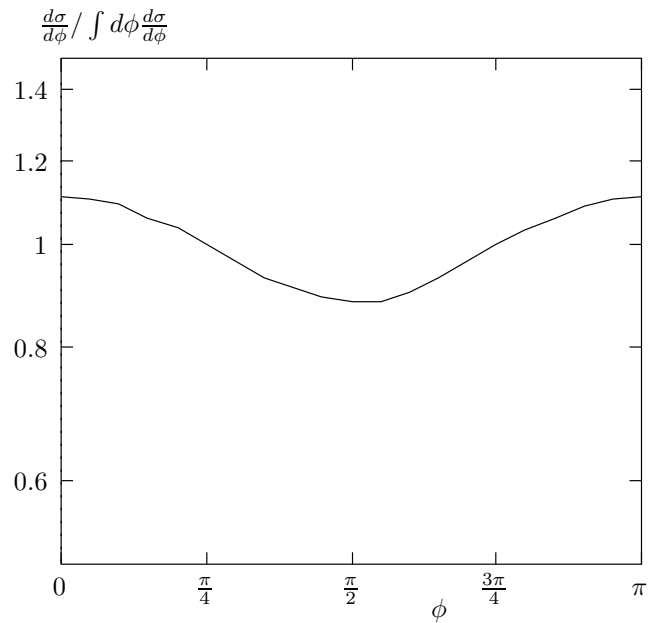
\begin{figure}[htbp]
\begin{center}
\input fig3.pstex_t
\end{center}
\caption{
\label{f5}}
\end{figure}
\begin{figure}[htbp]
\begin{center}
\input fig4.pstex_t
\end{center}
\caption{
\label{f6}}
\end{figure}
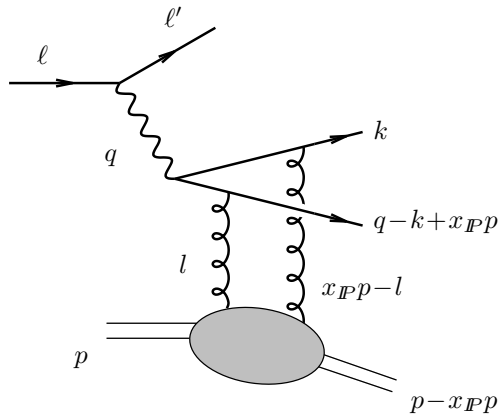
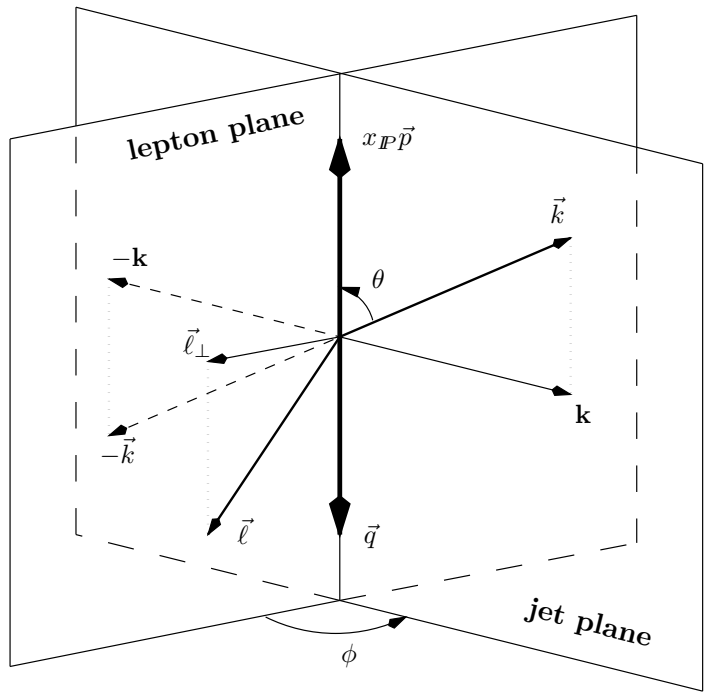
\end{document}